\documentclass[preprint2]{aastex6} 

\begin{document}

\newcommand{\kms}        {km~s$^{-1}$}
\newcommand{\herschel} {\textit{Herschel}}
\newcommand{\den}{${n}_{\mathrm{H}_2}$}
\newcommand{\ncolone}{$N_{\mathrm{CO}}$}
\newcommand{\tkin}{$T_{\mathrm{kin}}$}
\newcommand{\cmone}{cm$^{-1}$}
\newcommand{\cmtwo}{cm$^{-2}$}
\newcommand{\cmthree}{cm$^{-3}$}
\newcommand{\coone}{$^{12}$CO}
\newcommand{\msun}{M$_{\odot}$}
\newcommand{\lsun}{L$_{\odot}$}
\newcommand{\mic}{$\mu$m}
\newcommand{\htwo}{H$_{2}$}
\newcommand{\lfir} {$L_{\mathrm{FIR}}$}
\newcommand{\lco} {$L_{\mathrm{CO}}$}
\newcommand{\acet} {$\mathrm{C}_{2}\mathrm{H}_{2}$}
\newcommand{\xmol} {$\mathrm{C}_{3}\mathrm{H}_{3}^{+}$}
\newcommand{\ymol} {c-$\mathrm{C}_{3}\mathrm{H}_{2}$}
\newcommand{\acetiso} {$^{13}\mathrm{C}\mathrm{C}\mathrm{H}_{2}$}
\newcommand{\vlsr}{V$_\mathrm{LSR}$}
\newcommand{\vfwhm}{V$_\mathrm{FWHM}$}
\newcommand{\cratio}{$^{12}\mathrm{C}/^{13}\mathrm{C}$}
\newcommand{\amon}{NH$_{3}$}

\def\msol{\ifmmode {\>M_\odot}\else {$M_\odot$}\fi}
\def\lsol{\ifmmode {\>L_\odot}\else {$L_\odot$}\fi}

\def\be{\begin{equation}}
\def\ee{\end{equation}}
\def\bdm{\begin{displaymath}}
\def\edm{\end{displaymath}}
\def\bea{\begin{eqnarray}}
\def\eea{\end{eqnarray}}

\title{High Spectral Resolution SOFIA/EXES Observations of \acet\ towards Orion-IRc2} 

\author{Naseem Rangwala\altaffilmark{1,2}, Sean W. J. Colgan\altaffilmark{1}, Romane Le Gal\altaffilmark{8}, Kinsuk Acharyya\altaffilmark{5}, Xinchuan Huang\altaffilmark{1}, Timothy J. Lee\altaffilmark{1}, Eric Herbst\altaffilmark{4}, Curtis deWitt\altaffilmark{6}, Matt Richter\altaffilmark{6}, Adwin Boogert\altaffilmark{3,7}, and Mark McKelvey\altaffilmark{1}}

\altaffiltext{1}{Space Science and Astrobiology Division, NASA Ames Research Center, Moffet Field, CA 94035}  
\altaffiltext{2}{Bay Area Environmental Research Institute, Moffet Field, CA 94035}
  \altaffiltext{3}{Universities Space Research Association, SOFIA Science Center, Moffet Field, CA 94035}
  \altaffiltext{4}{Departments of  Chemistry and Astronomy, University of Virginia, McCormick Rd, Charlottesville, VA 22904}
   \altaffiltext{5}{Physical Research Laboratory, Ahmedabad, India}
   \altaffiltext{6}{University of California, Davis, Phys 539, Davis, CA 95616}
   \altaffiltext{7}{Institute for Astronomy, University of Hawaii, 2680 Woodlawn Drive,Honolulu, HI 96822}
   \altaffiltext{8}{Harvard-Smithsonian Center for Astrophysics, 60 Garden St., Cambridge, MA 02138}
  
\begin{abstract}
We present high-spectral resolution observations from 12.96 - 13.33 microns towards Orion IRc2 using the mid-infrared spectrograph, EXES, on SOFIA. These observations probe the physical and chemical conditions of the Orion hot core, which is sampled by a bright, compact, mid-infrared background continuum source in the region, IRc2. All ten of the rovibrational \acet\ transitions expected in our spectral coverage, are detected with high S/N, yielding continuous coverage of the R-branch lines from J=9-8 to J=18-17, including both ortho and para species. Eight of these rovibrational transitions are newly reported  detections. The isotopologue, \acetiso, is clearly detected with high signal-to-noise. This enabled a direct measurement of the \cratio\ isotopic ratio for the Orion hot core of $14 \pm 1$ and an estimated maximum value of 21. We also detected several HCN rovibrational lines. The ortho and para \acet\ ladders are clearly separate and tracing two different temperatures, 226~K and 164~K, respectively, with a non-equilibrium ortho to para ratio (OPR) of $1.7 \pm 0.1$. Additionally, the ortho and para \vlsr\ values differ by about $1.8 \pm 0.2$ \kms, while, the mean line widths differ by $0.7 \pm 0.2$ \kms, suggesting that these species are not uniformly mixed along the line of sight to IRc2. We propose that the abnormally low \acet\ OPR could be a remnant from an earlier, colder phase, before the density enhancement (now the hot core) was impacted by shocks generated from an explosive event 500 yrs ago.

\end{abstract}


\section{Introduction}

Acetylene (\acet), being one of the simpler and abundant hydrocarbons, is thought to play a major role in interstellar chemistry. It is an important building block in the gas phase formation of larger hydrocarbons, ring molecules and PAHs in the interstellar medium \cite[][e.g.,]{winnewisser79, schiff79, herbst83, millar84, contreras2013}. It is also an important precursor in the formation of nitriles or cyanopolyynes in the ISM, especially in hot molecular cores \cite[e.g.,][]{chapman2009, balucani2000}. Recent laboratory experiments show that \acet\ could have a significant impact on the growth of dust grains in carbon rich environments such as outflows from carbon stars, resulting in larger and more spherical grains (Sciamma-O'Brien et al.\,{\it in preparation}). 

Because \acet\ has no permanent dipole moment, it cannot be observed via rotational transitions at radio wavelengths, like CO or HCN.  Hence, \acet\ can only be studied in the mid-infrared (MIR; 5 - 28 \mic), where its rovibrational transitions lie. Its $\nu_{5}$ vibration-rotation band at 13.7 \mic\ is the strongest,  consisting of Q ($J_\mathrm{u} = J_\mathrm{l}$), R ($J_\mathrm{u} = J_\mathrm{l}+1$), and P ($J_\mathrm{u} = J_\mathrm{l}-1$) branch lines. The 3:1 statistical weights arising from the degeneracy associated with the nuclear spin of identical {\bf protons} favor the odd-J lower state transitions. 

The first astronomical detection of \acet\ was by \citet{lacy89} in four sources, including Orion IRc2.  These measurements utilized an echelle spectrograph on NASA's infrared telescope facility (IRTF) 3m telescope on Mauna Kea, with a spectral resolution of 25 \kms (R = 15000). They detected unresolved R(5), R(10), and Q-branch lines towards IRc2. Later, \citet{evans91} used the same telescope and instrument to obtain more transitions of \acet. More recently, the TEXES mid-IR spectrograph became available on NASA IRTF \citep{lacy2002}, providing a significant improvement in data quality and spectral resolution. Sample observations from this instrument are shown in \citet{lacy2002}, including the measurement of one \acet\ transition in IRc2. More \acet\ lines were observed using TEXES towards IRc2 (M. Richter; Private Communication), but the data have not been published. 

\acet\ has also been observed from space towards many sources in the ISM, including Orion IRc2 \citep{vandishoeck98, boonman2003} using telescopes such as \textit{ISO} and \textit{Spitzer}. However, the space-based astronomical IR/MIR missions have insufficient spectral resolution to resolve the individual rovibrational transitions that are needed to measure physical conditions and accurate column densities. These observations are only able to link observed bands to the most abundant molecules, without actually identifying specific molecular transitions. 

The Stratospheric Observatory for Infrared Astronomy (SOFIA) is a Boeing 747 carrying a 2.5m telescope up to 45,000 feet, above 99\% of the earth's water vapor, enabling astronomical observations from near-IR to far-IR wavelengths. The Echelon-Cross-Echelle Spectrograph \citep[EXES;][]{richter2010} on SOFIA provides high spectral resolution in the MIR and is well suited to observe around 13.7~\mic\ --  the center of the \acet\ band. Additionally, SOFIA's beam size {\bf($3.2\arcsec \times 3.2\arcsec$)} being much smaller compared to space based missions like ISO ($14\arcsec \times 27\arcsec$), is able to isolate the source emission from the surrounding emission. Even though the atmospheric transmission from the ground at $\sim$13~\mic\ is reasonably good, the difference between a site like Mauna Kea at 14,000 feet and SOFIA gets increasingly large towards the center of the \acet\ 13.7~\mic\ band. In addition, the atmospheric lines at sea level or at 14,000 feet are significantly pressure broadened, which can also substantially impact line measurements. EXES is a sister instrument of TEXES so they are similar in many ways, although the spectral coverage of TEXES at a given wavelength is smaller than EXES ($256^2$ vs $1024^2$).

%Since, the primary goal of our observations from SOFIA was to search for \xmol, which is expected to have much weaker line signature, we proposed to observe with SOFIA to avoid any extra interference from the atmosphere and line broadening. 

\begin{figure}[ht]\label{irimage}
\includegraphics[scale=0.27]{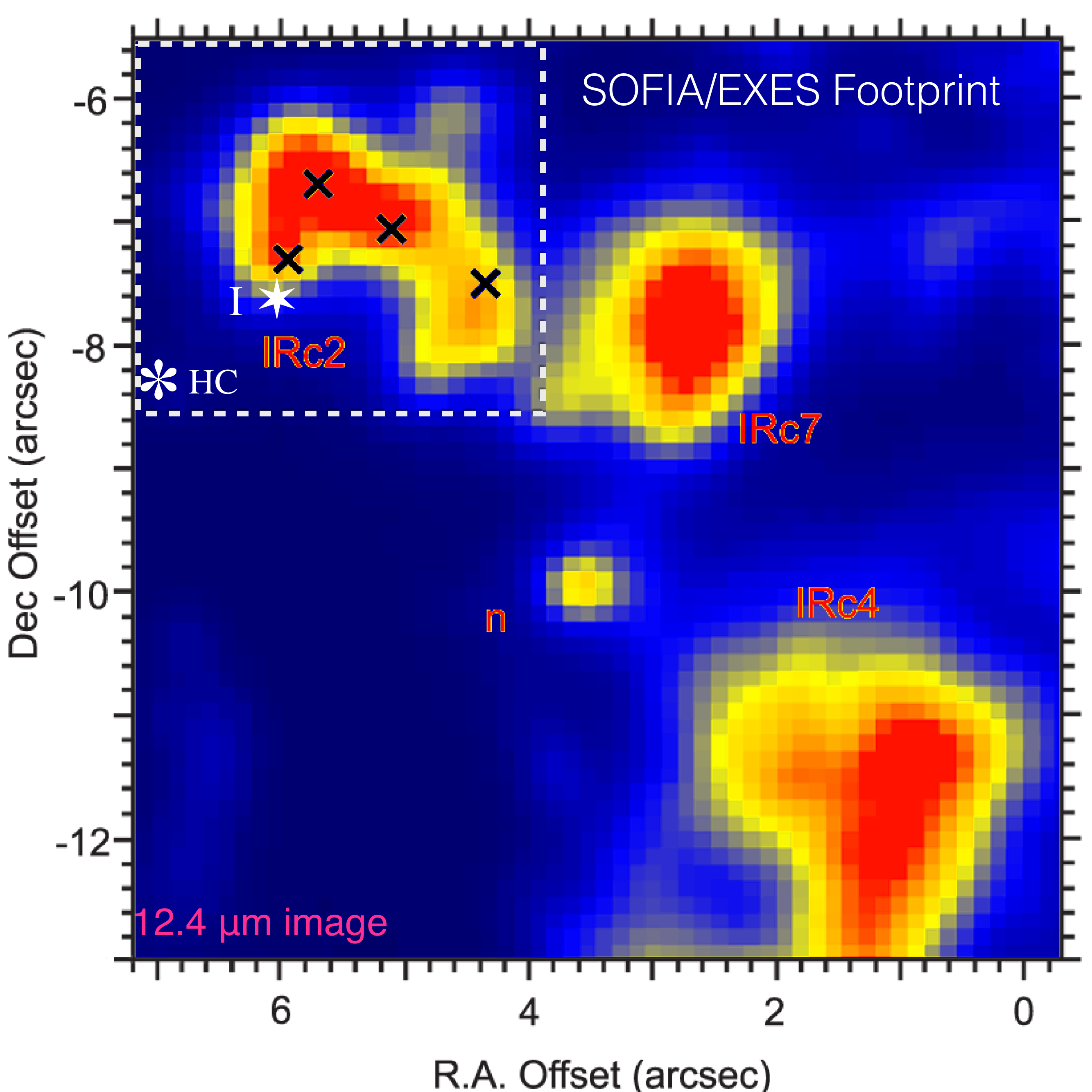}
\caption{Mid-infrared 12.4~\mic\ image of a portion of the Orion-KL region surrounding IRc2 from \citet{okumura2011}. The white asterisk and star are the locations of the hot core and radio source I. The dashed box is the 3\arcsec $\times$ 3\arcsec\ EXES footprint covering IRc2, which is a bright IR source closest to the hot core.}
\label{okumura}
\end{figure}

The Orion BN-KL region is the nearest (418~pc) and best-studied region of massive star formation \citep[e.g.,][]{bally2008, genzel89}. It contains a cluster of heavily dust-obscured infrared compact sources, including three radio sources thought to contain massive young stellar objects ($\sim$10 -- 20 \msun): the Becklin-Neugebauer (BN) object, and radio source I \citep{menten95} and IR source\textit{n} \cite{lonsdale82}. The BN object is the brightest MIR source in this region. Source I is a highly embedded radio source and Source \textit{n} is thought to be a relatively evolved young stellar object. The location of source I, n and the hot core relative to some of the brightest IR sources in the region are shown in Figure \ref{irimage}. This region also contains a spectacular wide angle outflow covering the entire Orion-KL region seen both in molecular and ionized gas \citep{bally2015,bally2017} caused by an explosion roughly 500 yrs ago, when BN and source I were ejected from the center. Millimeter observations \citep{beuther2005, plambeck1982} revealed the existence of a hot core, with an estimated temperature between 100 -- 350~K, located 1.2\arcsec\ southeast of source I. Hot molecular cores ($\sim$10000 AU) are characterized by high temperatures ($\gtrsim 100$~K) and densities ($\gtrsim 10^{6}$~\cmthree) with a large abundance of both simple and complex organic molecules \citep[e.g.,][]{Herbst09review}. 
No MIR or radio counterpart to the Orion hot core has been detected, demonstrating that this object is not heated internally. The Orion hot core has one of the richest molecular chemistries associated with massive star formation \citep[e.g.,][] {feng2016}. However, most of the high-spectral resolution studies of its chemistry have been done in the radio, (sub-) mm, and FIR wavelength regimes. In the MIR, the hot core composition is best sampled by the closest, bright  continuum source in this region, IRc2, located roughly 2\arcsec\ NW of the hot core peak \citep[e.g.,][]{okumura2011}. The IRc2 continuum can be used to sample the molecular material in front, assuming it is behind the hot core, to probe its chemistry, kinematics, geometry and physical conditions. The Orion hot core is extended $\sim 15\arcsec \times 15\arcsec$ \citep[from ammonia maps in][]{goddi2011} relative to the SOFIA/EXES beam.

In this paper, we present an EXES $\sim 5$ \kms\ resolution spectrum of Orion IRc2 from 12.96 -- 13.33~\mic.  These observations have better spectral resolution, higher S/N and continuous coverage of new \acet\ and HCN rovibrational transitions compared to previous published observations. The primary goal for these observations was to obtain a first detection of \xmol, which is considered to be the most important precursor in the formation of \ymol. We present motivation for the search of this molecule and the first measurement of an upper limit on its column density from these observations. Comparison of all these observations with astrochemistry models is also presented. 

\begin{figure*}[ht]
\includegraphics[scale=0.5]{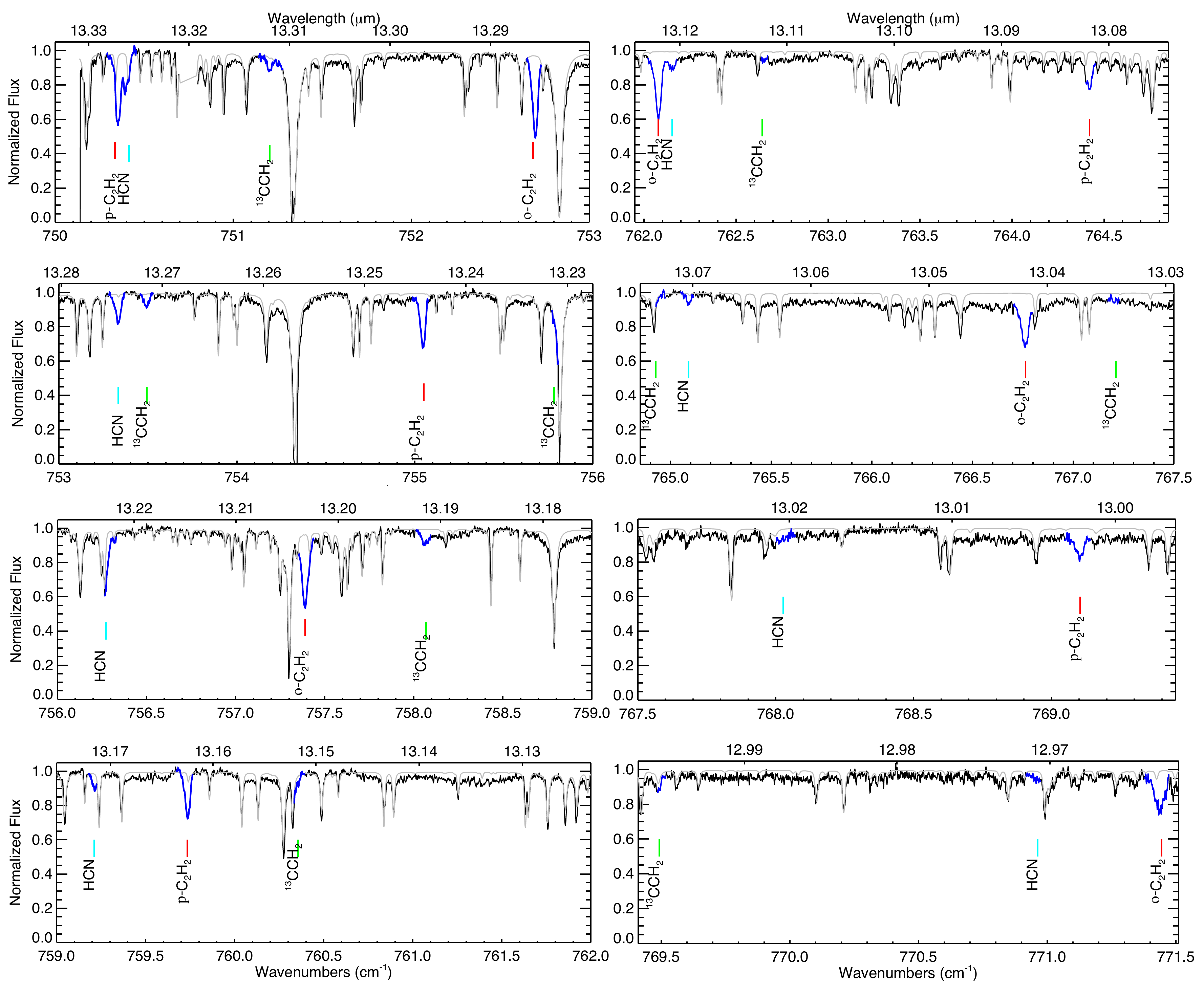}
\caption{EXES spectrum of Orion IRC2 from 750.15 -- 771.5 \cmone\ in black showing rovibrational transitions of \acet\ and HCN. An atmospheric model is overplotted in grey. Molecular lines from the astronomical source are highlighted in blue. \acet, \acetiso, and HCN transitions are labelled with red, green, and cyan lines, respectively.}
\label{spec}
\end{figure*}

\section{Observations and Data Reduction}
Orion IRc2 was observed by the EXES instrument (program ID 03\textunderscore0126) onboard SOFIA on 2015, October 3 at an altitude of 43000ft. Spectra were acquired in the cross-dispersed high-resolution mode over 32 orders to provide coverage from 12.96~\mic\ to 13.33~\mic. The slit was essentially a 3.2\arcsec $\times 3.2$\arcsec square, providing a resolving power of about 60,000 ($\sim$5 \kms), with the resolution element sampled by 16 pixels. The slit size covers the entire emission from IRc2. The total on-source integration time was 494 seconds giving an RMS ~0.017 erg s$^{-1}$ cm$^{-2}$ sr$^{-1}$ (cm$^{-1}$)$^{-1}$. The telescope was nodded in ABBA fashion every 34 seconds to enable the subtraction of telluric emission lines. The bright asteroid Vesta was observed about 2 hrs after the science observations at the same altitude. The on-source time on Vesta was not long and the resulting spectrum was noisier than the data from the science target. Hence, we do not use Vesta's spectrum to normalize the Orion IRc2's spectrum. Instead we use an atmospheric model spectrum generated by ATRAN to remove the atmospheric contribution and separate the atmospheric lines from the astrophysical lines. 

The pipeline first linearizes and calculates the response for each pixel. Prior to the observation of the target source, we took a featureless spectrum by rotating a folding mirror which directs a 260 +/- 0.1 K blackbody source into the beam path. This blackbody spectrum is used to locate the echelon order edges for the purposes of calculating the distortion correction, removing the spectrograph blaze function and removing the pixel-to-pixel detector variations.
The on-slit and off-slit observations of the target were differenced to remove emission sky lines and telescope thermal emission. At the same time, we process a combined A+B spectrum, which is dominated by sky lines, for the purpose of determining the dispersion solution. 
The A-B target spectrum was divided by the normalized flat field and then rectified. The rectified orders were extracted using optimal weighting according to the spatial profile of the object. Lastly, the dispersion solution was calculated using the positions of known sky emission lines present in the rectified, extracted sky line spectrum. 

\begin{figure*}[ht]
\includegraphics[scale=0.47]{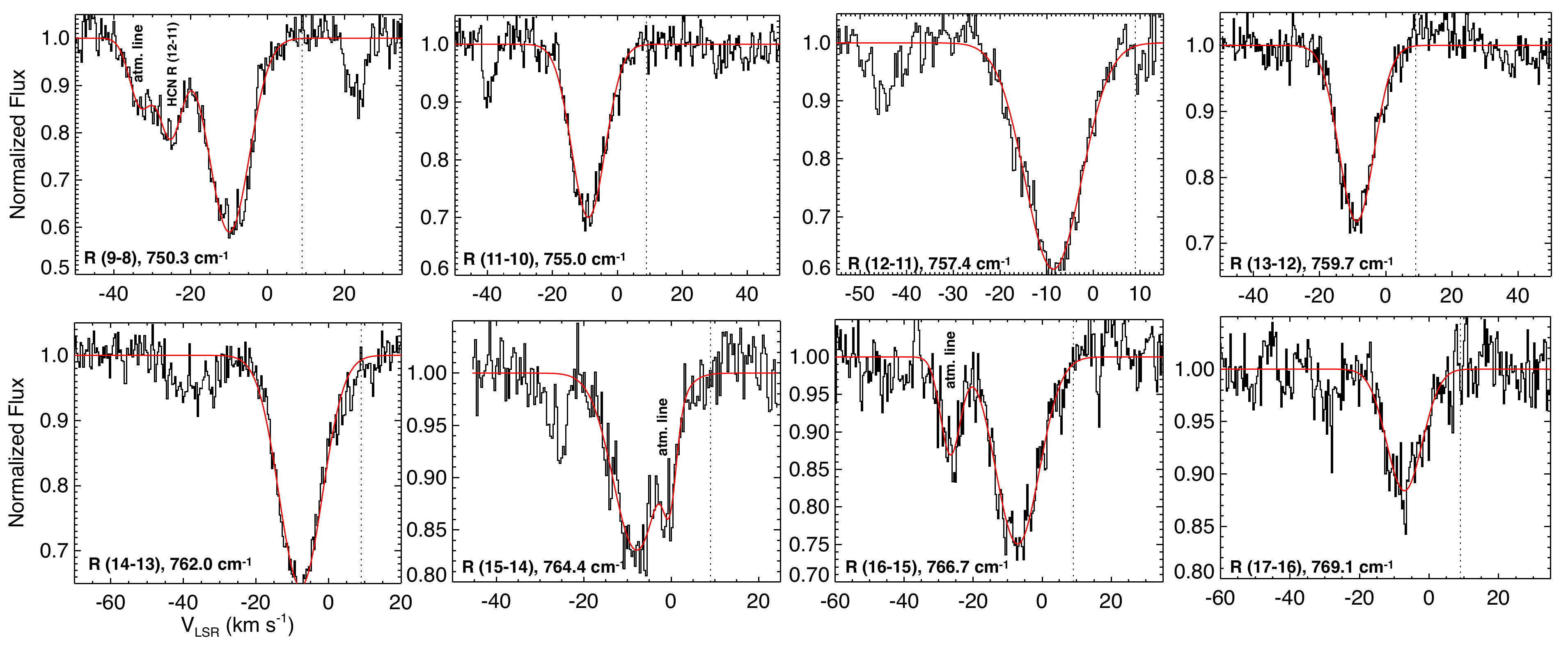}
\caption{Gaussian fits for select \acet\ lines are shown. In cases where multiple gaussians were fit, the additional features are labeled. The systemic velocity of the ambient cloud is $\sim$9 \kms \citep{genzel89}. }
\label{fits}
\vspace*{0.08in}
\end{figure*}

We use the merged spectrum made from the individual orders of the echelon. This wavelength calibration is accurate to within $\pm0.1$ \kms.  Figure \ref{spec} shows the observed EXES spectrum of Orion IRc2 in black. The model atmospheric transmission from ATRAN is plotted in gray. All of the ten rovibrational transitions of \acet, expected in our spectral coverage, are detected, yielding continuous coverage of the R-branch lines from J = 9 -- 8 to J = 18 -- 17, which include both ortho (odd J lower states) and para (even J) species. These lines are marked in red. The isotopologue, \acetiso, is clearly detected with high signal-to-noise. Three transitions of \acetiso\ are well separated from the atmospheric lines, the rest are either very low-S/N or blended with atmospheric lines. Six rovibrational transitions from the $\nu_2$ band of HCN are also detected. The expected line positions of \acetiso\ and HCN are labelled with green and cyan lines, respectively. Select spectra for individual \acet\ lines of ortho and para species are shown as a function of \vlsr\ in Figure \ref{fits}. 

\section{Analysis}

The molecular transitions in Figure \ref{spec} do not exhibit normalized fluxes which drop below $\sim0.65$. Compared with NGC 7538 IRS1 \citep{knez2009}, the \acet\ features are broader, deeper, and show a much greater range of apparent absorption. Compared with previous \acet\ observations of IRc2, more lines and a greater variety of ortho and para transitions are measured with improved spectral resolution. For our initial analysis, we assume optically thin lines from absorbing material which fully covers the background continuum source. Given the size of SOFIA beam and amount of emission in the BN/KL region, we then consider partial filling factors by optically thick material. 

\subsection{Line Fitting}
The absorption lines from the source, as highlighted in Figures \ref{spec} and \ref{fits}, were fit assuming a Gaussian line profile, using the following function taken from \citet{indriolo2015}:
\begin{eqnarray}\label{gauss}
I &=& I_{0}\large[1 - f_\mathrm{c}(1 - e^{-\tau_{0}G})], \,\,\, \mathrm{where} \\
G &=& \mathrm{exp}\big[-\frac{(v - v_\mathrm{c})^{2}}{2\sigma^{2}_{v}}]\nonumber
\end{eqnarray}
$I_{0}$ is the continuum level, $\tau_{0}$ is the line center optical depth, $v_\mathrm{c}$ is the velocity at the line center and $\sigma_{v}$ is the velocity dispersion. A covering factor, $f_\mathrm{c}$ to account for only partial coverage by the absorbing material, is also included. 

Prior to fitting, the absorption lines were normalized by the average continuum level estimated by excluding the strong features from the vicinity of these lines. The line center, FWHM, and $\tau_0$ of the absorption lines were determined by fitting equation~\ref{gauss} using the MPFIT package in IDL. A linear or quadratic polynomial function was sufficient to fit the continuum baseline. Table 1 lists the best fit line parameters and their fitted 1-$\sigma$ errors for $f_\mathrm{c} =1$ and 0.5, along with the relevant molecular parameters (HITRAN; Rothman et al.\,2012). The column density in the lower state of the observed transition is estimated by integrating over the absorption line using 
\begin{equation}
dN/dv = \tau(v) \frac{g_{l}}{g_{u}}\frac{8\pi}{A\lambda^{3}}.
\end{equation}
\\
The constants $g_{l}$ and $g_{u}$ are statistical weights in the lower and upper states, respectively, $A$ is the Einstein coefficient for spontaneous emission and $\lambda$ is the central wavelength of a given transition. We first consider $f_\mathrm{c} =1$, discussion for a lower covering factor is presented later. 

\subsection{Rotation Diagrams}
\begin{figure*}
\includegraphics[scale=0.64]{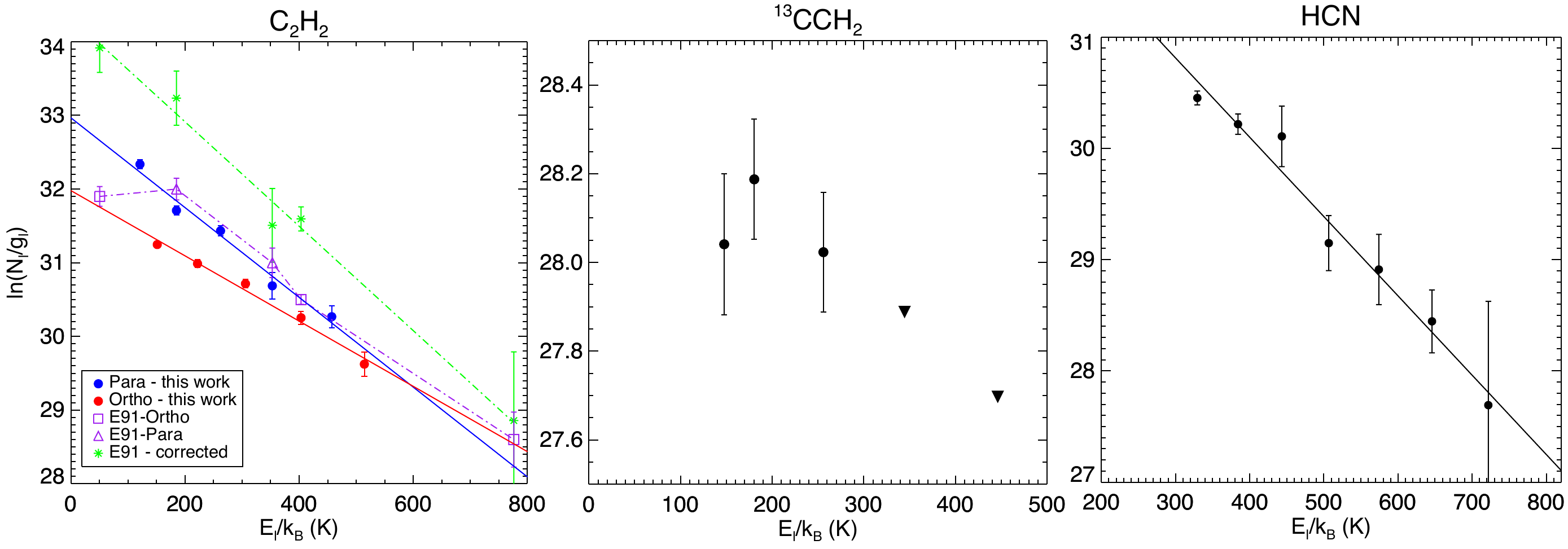}
\caption{Left panel: Rotational diagrams for \acet. Comparison of \acet\ with \citet{evans91} is included in the left panel for their uncorrected (purple dashed line) and corrected (green dashed line, see text) data. Red, blue and green lines are linear fits to the data. Middle panel: Rotation diagram for \acetiso. Downward triangles show $3-\sigma$ upper limits. No physical fit was found for the \acetiso\ transitions. Right panel: Rotation diagram for HCN. Black line is a linear fit to the HCN data.}
\label{comp}
\end{figure*}

In Figure \ref{comp}, we plot the natural log of column density per statistical weight for each transition as a function of the energy level of the lower state, since only absorption lines were found. This is usually referred as a rotation or population diagram \citep{goldsmith99}. In local thermodynamic equilibrium (LTE), the molecules follow a Boltzmann distribution and the points in the rotation diagram can be fit using:  

\begin{equation}\label{BD}
\textrm{ln}\frac{N_{j}}{g_{j}} = \textrm{ln}\frac{N}{Q_R (T_\textrm{ex})} - \frac{E_l}{kT_\textrm{ex}},
\end{equation}

\begin{deluxetable*}{lllllllll}
\tablecaption{Observed \acet\ ($\nu_\mathrm{4}$ band) ro-vibrational Transitions and Inferred Parameters}
\tablecolumns{9}
\tablenum{1}
%\tablehead{\colhead{Transition} & \colhead{Wavenumber} &  \colhead{E$_{\u}$/k$_{b}$} & \colhead{g$_{u}$} & \colhead{$A$} & \colhead{v_{LSR}} & \colhead{v$_{FWHM}$} & \colhead{EW}}
\tablehead{
\colhead{Transition} & 
\colhead{Wavenumber} & 
\colhead{$E_\mathrm{l}/k_\mathrm{b}$} & 
\colhead{$g_\mathrm{l}$} & 
\colhead{$A$} & 
\colhead{$V_\mathrm{LSR}$} & 
\colhead{$V_\mathrm{FWHM}$} & 
\colhead{$\tau_{0}$} &
\colhead{$N_\mathrm{l}$} \\
\colhead{($J^{\prime} - J^{\prime\prime}$)} & \colhead{(cm$^{-1}$)} & \colhead{(K)} & \colhead{} & \colhead{(s$^{-1}$)}  & \colhead{(\kms)} & \colhead{(\kms)} & \colhead{} & \colhead{($\times 10^{15}$ \cmtwo)}
}
\startdata
%\cutinhead{$^{12}C_{2}H_{2}$ Para Series}
\cutinhead{$^{12}C_{2}H_{2}$ Para Series}
9-8 & 750.31057 & 121.0 & 17 & 3.454 & $-10.0 \pm 0.2$ & $11.9 \pm 0.4$ &  $0.54 \pm 0.02$ & $1.88 \pm 0.11$ \\
& $f_\mathrm{c} = 0.5$ & & & & & $9.8 \pm 0.3$ &  $1.54 \pm 0.08$ & $4.40 \pm 0.38$ \\
11-10 & 755.00459 & 184.9 & 21 & 3.488 & $-9.0 \pm 0.4$ & $11.6 \pm 0.4$ & $0.35 \pm 0.01$ &$1.24 \pm 0.08$ \\
& $f_\mathrm{c} = 0.5$ & & & & & $10.9 \pm 0.3$ & $0.88 \pm 0.03$ &$2.91 \pm 0.21$ \\
13-12 & 759.69522 & 262.2 & 25 & 3.533 & $-8.8 \pm 0.4$ & $11.9 \pm 0.4$ & $0.31 \pm 0.01$  &$1.12 \pm 0.08$\\
& $f_\mathrm{c} = 0.5$ & & & & & $11.4 \pm 0.4$ & $0.74 \pm 0.03$  &$2.57 \pm 0.20$\\
15-14 & 764.38213 & 352.9 & 29 & 3.586 & $-7.9 \pm 0.5$ & $10.8 \pm 1.2$ & $0.18 \pm 0.01$  &$0.62 \pm 0.11$\\
& $f_\mathrm{c} = 0.5$ & & & & & $10.4 \pm 1.2$ & $0.41 \pm 0.03$  &$1.31 \pm 0.24$\\
17-16 & 769.06500 & 457.1 & 33 & 3.646 & $-7.1 \pm 0.9$ & $11.9 \pm 0.9$ & $0.12 \pm 0.01$  &$0.46 \pm 0.07$\\
& $f_\mathrm{c} = 0.5$ & & & & & $11.6 \pm 0.9$ & $0.26 \pm 0.02$  &$0.96 \pm 0.15$\\
\cutinhead{$^{12}C_{2}H_{2}$ Ortho Series}
10-9 & 752.65799 & 151.3 & 57 & 3.47 & -7.6 $\pm$ 0.3 & $11.9 \pm 0.2$ & $0.60 \pm 0.01$ & $2.12 \pm 0.09$ \\
& $f_\mathrm{c} = 0.5$ & & & & & $10.5 \pm 0.2$ & $1.95 \pm 0.09$ & $6.09 \pm 0.43$ \\
12-11 & 757.35035 & 221.9 & 69 & 3.509 & -8.6 $\pm$ 0.4 & $13.4 \pm 0.4$ & $0.49 \pm 0.01$ & $1.98 \pm 0.11$\\
& $f_\mathrm{c} = 0.5$ & & & & & $12.3 \pm 0.3$ & $1.37 \pm 0.06$ & $5.10 \pm 0.37$\\
14-13 & 762.03916 & 305.9 & 81 & 3.559 & -7.6 $\pm$ 0.4 & $13.2 \pm 0.4$ & $0.43 \pm 0.01$  &$1.77 \pm 0.10$\\
& $f_\mathrm{c} = 0.5$ & & & & & $12.3 \pm 0.4$ & $1.16 \pm 0.05$  &$4.41 \pm 0.32$\\
16-15 & 766.72409 & 403.4 & 93 & 3.615 & -7.2 $\pm$ 0.3 & $14.3 \pm 0.6$ & $0.29 \pm 0.01$ &$1.28 \pm 0.11$\\
& $f_\mathrm{c} = 0.5$ & & & & & $13.6 \pm 0.6$ & $0.68 \pm 0.03$ &$2.87 \pm 0.28$\\
18-17 & 771.40482 & 514.2 & 105 & 3.678 & -4.7 $\pm$ 1.2 & $13.3 \pm 1.1$ & $0.18 \pm 0.01$ &$0.77 \pm 0.13$\\
& $f_\mathrm{c} = 0.5$ & & & & & $13.2 \pm 1.1$ & $0.41 \pm 0.04$ &$1.72 \pm 0.29$\\
\cutinhead{\acetiso}
10-9 & 751.16424 & 147.7 & 152 & 3.057 & -8.2 $\pm$ 0.6 & $12.5 $ & $0.055 \pm 0.006$ & $0.233 \pm 0.04$\\
11-10 & 753.45375 & 180.5 & 168 & 3.071 & -7.3 $\pm$ 0.6 & $12.5$ & $0.069 \pm 0.007$ & $0.293 \pm 0.04$ \\
13-12 &  758.03003 & 255.9 & 200 & 3.109 & -6.3 $\pm$ 0.9 & $12.5 $ & $0.068 \pm 0.007$ & $0.296 \pm 0.04$\\
\cutinhead{HCN}
12-11 & 750.35297	& 329.4	& 150 	& 1.242	& -9.3  $\pm$ 0.5 & $12.5$	& 		$0.240 \pm 0.011$ &		$2.52 \pm 0.16$ \\
 13-12 & 	753.29756 & 384.3	& 162	& 1.254	& -6.5 $\pm$ 0.3 & $14.0 \pm 0.7$ & 	$0.181 \pm	0.008$ &		$2.15	\pm 0.20$ \\
 14-13 & 756.24070 & 443.4	& 174	& 1.265	& -7.3 $\pm$ 1.0 & $12.0$	& 		$0.202 \pm	0.039$ &		$2.07	\pm 0.56$\\
 15-14 & 759.18245 & 506.7	& 186	& 1.278	& -4.8 $\pm$ 0.6 & $10.3 \pm 1.5$ & 	$0.095 \pm	0.009$ &		$0.85	\pm 0.21$\\
 16-15 & 762.12258 & 574.2	& 198	& 1.29	& -4.3 $\pm$ 1.0 & $13.4	\pm 2.2$ & $0.061 \pm 0.009$ &		$0.71	\pm 0.23$\\
 17-16 & 765.06107 & 645.9	& 210	& 1.302	& -4.0 $\pm$ 0.6 & $8.7	\pm 1.3$ &  $0.062	\pm 0.008$ &		$0.47	\pm 0.13$\\
 18-17 & 767.99780 & 721.8	& 222	& 1.315	& -0.3 $\pm$ 2.2 & $9.8	\pm 5.6$ & $0.027	\pm 0.009$ &		$0.24	\pm 0.22$\\
 19-18 & 770.93000& 802.0      & 234        & 1.327    & -- &  -- & -- & $< 0.6$ \\
\enddata
\tablecomments{When two rows are given, the first row is for $f_\mathrm{c} = 1.0$ and the second is for $f_\mathrm{c} = 0.5$.}
\tablecomments{There is an absolute additional error of 0.5 \kms\ when transforming from the geocentric frame (our data) to the LSR frame.}
 %Provide upper limits for lines not detected or obscured by atmosphere? 
\end{deluxetable*}

where $N$ is the total column density and $Q_R$ is the rotational partition function. Thus, a thermal distribution of populations will produce a straight line with a slope of 1/$T_\mathrm{ex}$ providing an estimate of the excitation temperature. $T_\mathrm{ex}$ is expected to be equal to the kinetic temperature if all levels were thermalized. The y-intercept of this line gives an estimate of the total column density ($N$) of the molecule if the rotational partition function, $Q_R$\footnote{For \acet, $Q_R = 0.0023\,T_\mathrm{ex}^{2} + 0.7014\,T_\mathrm{ex} + 12.858$, valid for $T_\mathrm{ex}$ up to 1000 K}, is known. For \acet, LTE is a good approximation -- because \acet\ has no permanent dipole moment, the collisions dominate over the radiative transitions between rotational levels.

\begin{figure*}[ht]
\center
\includegraphics[scale=0.44]{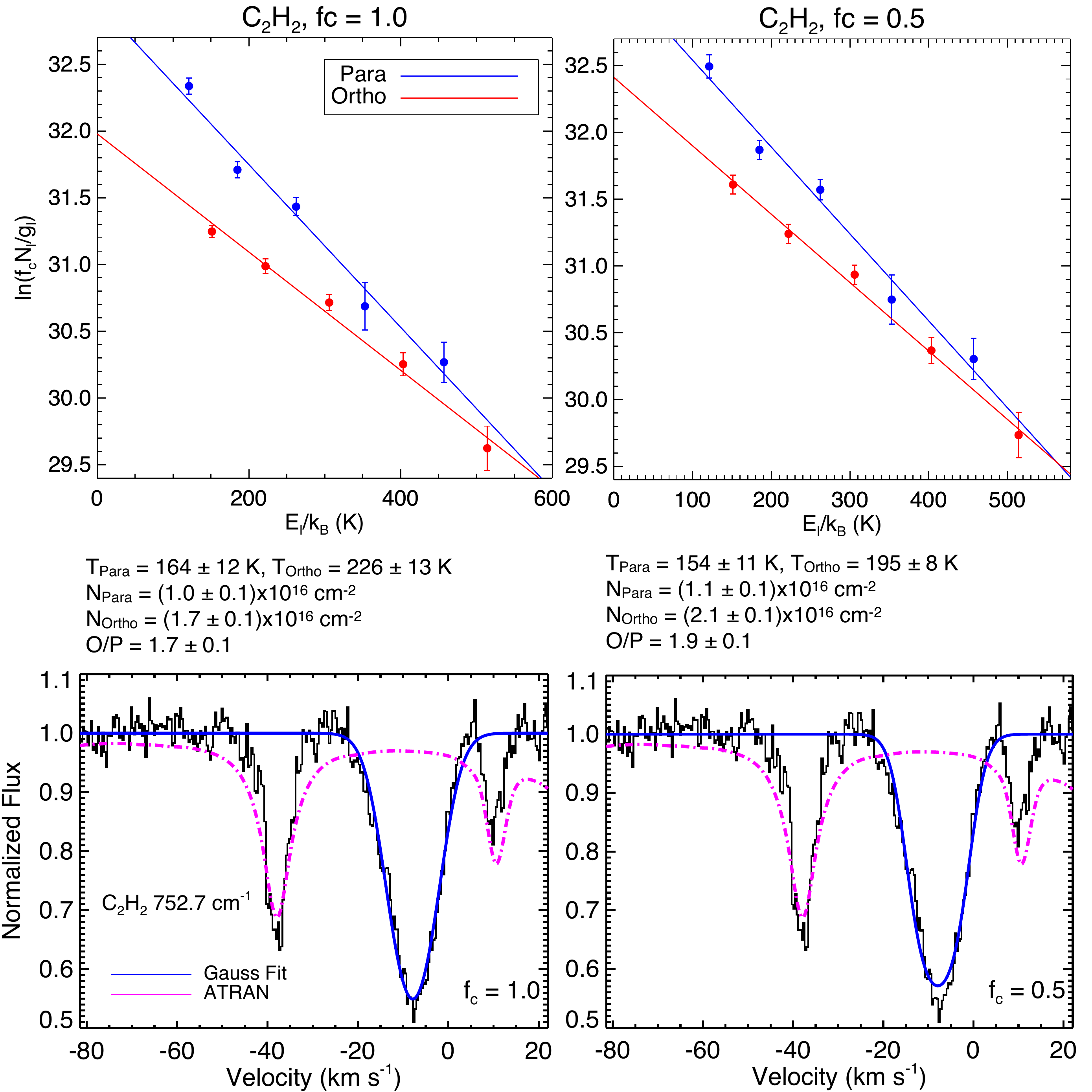}
\caption{Top panels: Rotational diagrams for ortho and para \acet\ towards Orion IRc2 are shown for two covering factors, 1.0 and 0.5. Bottom Panels: Effect on line fits for the two covering factors are shown. The ortho and para ladders are offset from each other and tracing two different temperatures. The offset is smaller but still present for coverage fraction of 0.5 .}
\label{rotdia}
\end{figure*}

In Figure \ref{comp} (left panel), the blue and red data points show the para and ortho ladders for \acet\ measured from the EXES data. The two ladders were fit independently using Equation 3. The total column density in ortho and para species was calculated using a separate partition function, $Q_R$, for ortho and para species, estimated using the fitted excitation temperatures and assuming that $Q_R$ (ortho) = 3/4 $Q_R$ (total) and $Q_R$ (para) = 1/4 $Q_R$ (total). For unity covering factor, the temperature and column density for the ortho \acet\ are $226 \pm 13$~K and {\bf ($1.7 \pm 0.1) \times 10^{16}$~\cmtwo}, respectively. The temperature and column density for the para \acet\ are $164 \pm 12$~K and {\bf($1.0 \pm 0.1) \times 10^{16}$~\cmtwo}, respectively. Transitions from the isotopologue, \acetiso, are expected to be much weaker, assuming the observed \cratio\ of $\sim$45 in our Galaxy. Three transitions of \acetiso\ are clearly detected, but with much lower S/N compared to \acet. The rest are either completely blended with atmospheric lines or below the detection limit. The rotation diagram for \acetiso\ is shown in Figure \ref{comp} (middle panel), including upper limits for transitions that do not have contamination from atmospheric lines in their vicinity. Because of fewer lines and large uncertainties, we are not able to obtain a physical fit to these data. Therefore, we use the average temperature from the ortho and para \acet\ analysis to estimate the total weighted column density in \acetiso\ of $(3.9 \pm 0.3)  \times 10^{15}$~\cmtwo, where the quoted uncertainty includes only the measurement errors and does not include uncertainties in the assumed temperature. For unity covering factor, we derive \cratio\ isotopic ratio of $14 \pm 1$ in the hot core. Fitting the HCN rotation ladder shown in Figure \ref{comp} (right panel), we get a temperature and total column density of $140 \pm 10$~K and ($8.4 \pm 0.6) \times 10^{16}$~\cmtwo, respectively.

In addition to ortho and para \acet\ showing distinct ladders on the rotation diagram, the two also show small but significant differences in their line centers and FWHM. The weighted mean \vlsr\ for ortho \acet\ ($-7.6 \pm 0.2$ \kms) is systematically higher than for para \acet\ ($-9.4 \pm 0.2$ \kms) by $1.8 \pm 0.2$ \kms. The weighted mean ortho line width ($12.5 \pm 0.16$ \kms) is wider than the mean para line width ($11.8 \pm 0.2$ \kms) by $0.7 \pm 0.2$ \kms. The weighted mean \vlsr\ of \acetiso\ is $-7.5 \pm 0.4$ \kms, consistent only with the ortho transitions. For HCN, the weighted mean \vlsr\ and mean line width are $-6.4 \pm 0.2$ \kms\ and $12.6 \pm 0.5$ \kms, respectively. 

The fitted parameters listed in Table 1 were obtained by fitting each line individually. Note that the \vlsr\ is not the same between the transitions; it gets systematically lower with $J$. This trend is true for both the ortho and para species, in addition to the average \vlsr\ being different between the ortho and para species. This trend would imply that the individual \acet\ transitions for the two species are coming from different locations with different temperatures. To check if this trend is real, we fit all the \acet\ transitions simultaneously, separately for the ortho and para species, with the same \vlsr\ and \vfwhm\ for each ladder. The resulting \vlsr\ and \vfwhm\  for the ortho  \acet\ are $-7.6 \pm 0.1$~\kms\ and $12.3 \pm 0.2$ \kms, respectively and for the para \acet\ are $-9.2 \pm 0.1$~\kms\ and $11.6 \pm 0.2$ \kms, respectively. There are no significant changes in the line center optical depths. The reduced $\chi^2$ for these fits are not significantly different from individual fits, implying that the \vlsr\ changing with J is not significant enough to conclude that the different transitions are spatially distinct.

Our observed \vlsr\ for \acet\ and HCN absorption lines are significantly blueshifted relative to the ambient molecular cloud velocity of 10 \kms. The same blueshifted absorption was also observed by Lacy et al. (2005) in \acet\ and HCN towards IRc2 using the TEXES instrument on the NASA IRTF telescope. In addition to absorption, they also mapped \acet\ in emission that is arising from the more diffuse material around the periphery of the Orion KL region. This emission was at the ambient cloud velocity. They concluded that this emission is associated with the gas that is outside the KL cavity. The absorption observed in our data and in Lacy et al.\,(2005) at the \vlsr $\sim$ -8 \kms\ is 18 km/s blueward of the molecular cloud. This value is similar to the approaching side of the master outflow originating within a few arcseconds of radio source I (e.g., Plambeck et al. 2009). It appears that the absorbing \acet\ and HCN are associated with this outflow.

We compare the lower level column densities of \acet\ transitions with previous observations of \citet[][E91 hereafter]{evans91}. In Figure \ref{comp} (left), we include the E91 data overplotted in purple; the squares are ortho \acet\ and the triangles are para \acet. Unlike our data, they do not have continuous coverage of the ortho and para lines to see the difference between the two ladders and they cannot spectrally resolve the lines. Thus, they fit both species with a single excitation temperature via equation \ref{BD} and find that their J = 5 level population (the leftmost square data point) is considerably lower than what would be predicted by any positive temperature. They attribute this behavior to saturation and correct for optical depth effects by fitting the absorption lines with a range of line widths. They find a best-fit velocity dispersion of 1.16 \kms, which was used to correct the rotation ladder to the values shown by the green data points in Figure \ref{comp}. Our data, which are fully resolved, agree well with their uncorrected data, but our line profiles disagree with a velocity dispersion of 1.16 \kms. This does not mean that the \acet\ lines are optically thin and free of saturation effects, but that the correction made by E91 does not agree with our more extensive measurements.

\subsection{\cratio\ Isotopic Ratio, Optical Depth Effects/Covering Factor and the Ortho-to-Para ratio }

The \cratio\ isotopic ratio can help identify abundance and optical depth effects. The observed \cratio\ absorption ratio is significantly greater than unity in our data, so that the \acetiso\ transitions are certainly optically thin. A comprehensive line survey towards Orion-KL \citep{tercero2010} reports a \cratio\ isotopic ratio of $\sim 45 \pm 20$. \citet{feng2016} also find a \cratio\ isotopic ratio that is consistent with \citep{tercero2010}, within their uncertainties. Compared with our derived isotopic estimate of $14 \pm 1$, this suggests the \acet\ lines are somewhat optically thick. For reference, the Galactic \cratio\ in our local ISM is $\sim$ 65 -- 75 \citep{wilson94, milam05}.

Our spectral resolution of 5 \kms\ fully resolves the \acet\ lines in Orion IRc2. The line profiles provide no evidence for saturation - the deepest line in our data only goes about 55\% below the continuum level. We also do not find any flat bottom line profiles - as one would expect from optically thick transitions. However, optically thick lines are still possible with such line shapes if the covering factor is $<1.0$, but the optical depth is not so high as to produce flat-bottomed profiles. From an examination of equation~\ref{gauss}, one notes that the observed maximum apparent absorption of $\sim$0.5 limits the value of $f_\mathrm{c}$ for optically thick material to $f_\mathrm{c} \gtrsim 0.5$. Although not completely relevant, unlike for NGC 7538 IRS1, the measured absorption dips for IRc2 vary by a factor of 2.5, consistent with these transitions being partially optically thin. 
The \acet\ spectra fits to equation~\ref{gauss} for $f_\mathrm{c} = 0.5$ are included in Table 1. The fit to the line profiles is not as good as for $f_\mathrm{c} = 1$, but still acceptable -- the change in $\chi ^{2}$ is not significant between the fits for $f_\mathrm{c} = 1$ and $f_\mathrm{c} = 0.5$.  The corresponding rotation diagram plots are presented in Figure \ref{rotdia} for $f_\mathrm{c} = $1 and 0.5. We show both the impact on the rotation diagram (top panel) and the line shape for one of the strongest \acet\ lines (bottom panel) in our data.  For $f_\mathrm{c} = 0.5$, the difference in the slope of the rotational diagram between the ortho and para ladders gets smaller than for $f_\mathrm{c} = $1.0, but there is still a significant difference in the excitation temperature. The $f_\mathrm{c} = 0.5$ total column density of ($3.2 \pm 0.1) \times 10^{16}$~\cmtwo\ implies a \cratio\ of $16.4 \pm 1.4$ with a 3$\sigma$ upper limit of 21.

In Figure \ref{comp}, the ortho and para ladders appear to be clearly separate and tracing two different temperatures. From the $f_\mathrm{c} = 1.0$ temperature and column density fits, we derive an ortho-to-para ratio (OPR) of $1.7 \pm 0.1$, much lower than the LTE value of 3. For $f_\mathrm{c} = 0.5$, the difference in excitation temperature between the ortho and para ladders is still present, with a small increase in the non-equilibrium OPR of 1.9. As noted in the previous section, the ortho and para \vlsr\ and line widths differ significantly, further supporting the idea that these species are not uniformly mixed along the line of sight to IRc2. 
\subsection{Comparison to Previous Work}
\begin{figure}[ht]
\center
\includegraphics[scale=0.25]{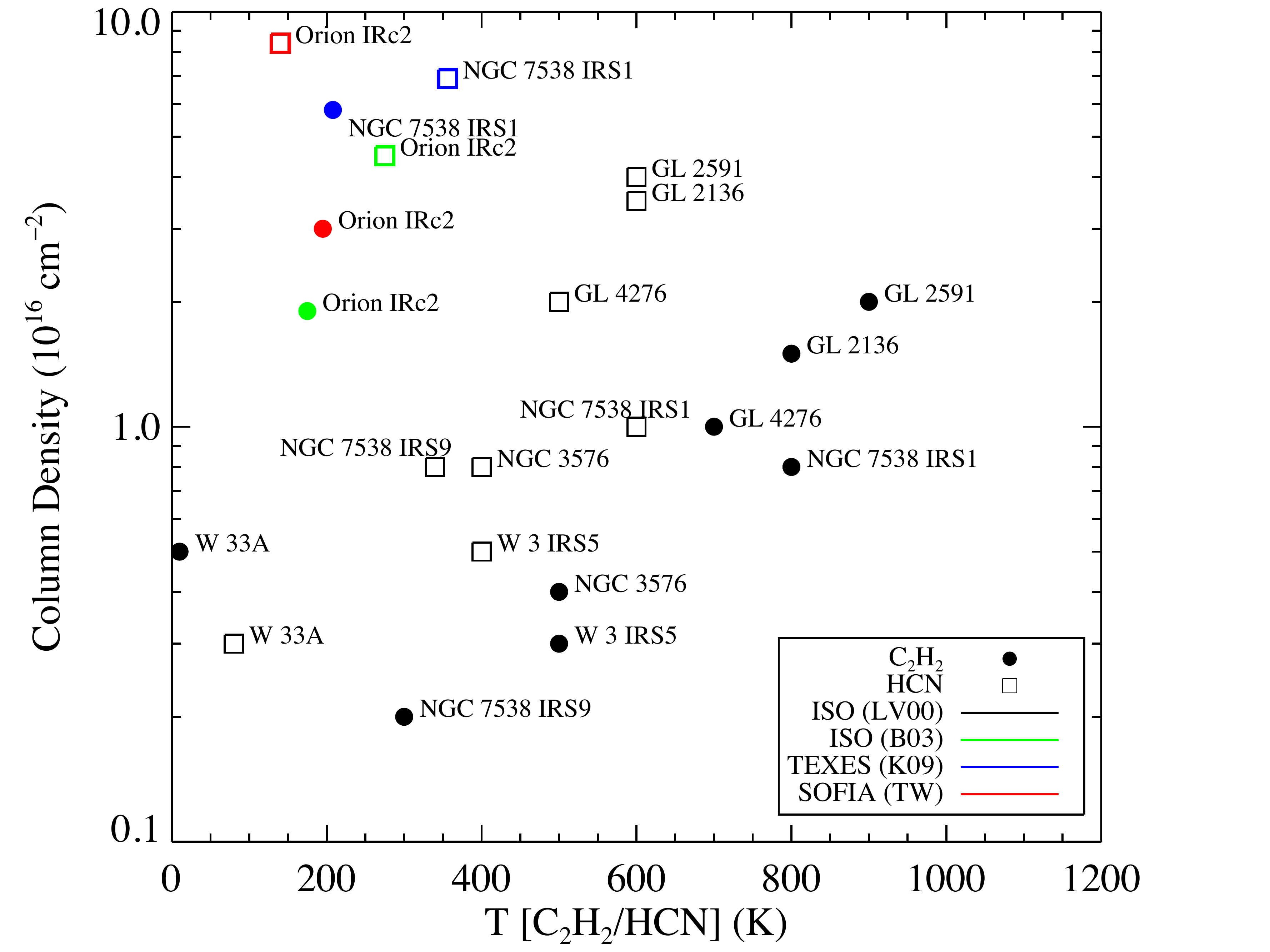}
\caption{Comparison to previous work showing column density versus excitation temperature for a variety of hot cores. Squares and filled circles are for HCN and \acet, respectively. All ISO data shown here come from \citet[][LV00]{lahuis2000} (black) except for Orion IRc2 (green), which comes from \citet[][B03]{boonman2003}. TEXES data for NGC 7538 IRS \citep[][K09]{knez2009} are shown in blue, while data from this work are shown in red.}
\label{comp_previous}
\end{figure}

In Figure \ref{comp_previous}, we compare high-resolution spectral and spatial measurements of \acet\ and HCN column densities with lower resolution ISO measurements of several hot cores/YSOs from \citet{lahuis2000} and for Orion IRc2 from \citet{boonman2003}. The high-resolution measurements for Orion IRc2 are from this work and for NGC 7538 IRS1 are from \citet{knez2009} using IRTF/TEXES. The present temperature and column density observations of Orion IRc2 differ from ISO's  for HCN and \acet. Most likely, this is because in ISO's bigger beam, the Orion IRc2's flux is dominated by the emission from the BN/KL region.  SOFIA measured lower temperatures and larger column densities for both HCN and \acet. The same trend is also seen for NGC 7532 IRS1, for which TEXES observations show large deviation from ISO's measurements. In this case, \citet{knez2009} argue that the combination of not accounting for saturation effects and larger line widths are responsible for the difference in column density and excitation temperature; the lower spectral resolution of ISO could not determine the level of saturation in the shallow lines observed by TEXES. In the light of these high-resolution observations, the conclusion from \citet{lahuis2000} that the column densities and abundances for the hot cores increase with increasing temperature, based on the trend in Figure \ref{comp_previous}, may no longer be true.

\section{Discussion}
\subsection{Comparison with Hot Core Models}

\begin{figure*}[ht]
\center
\includegraphics[scale=0.45]{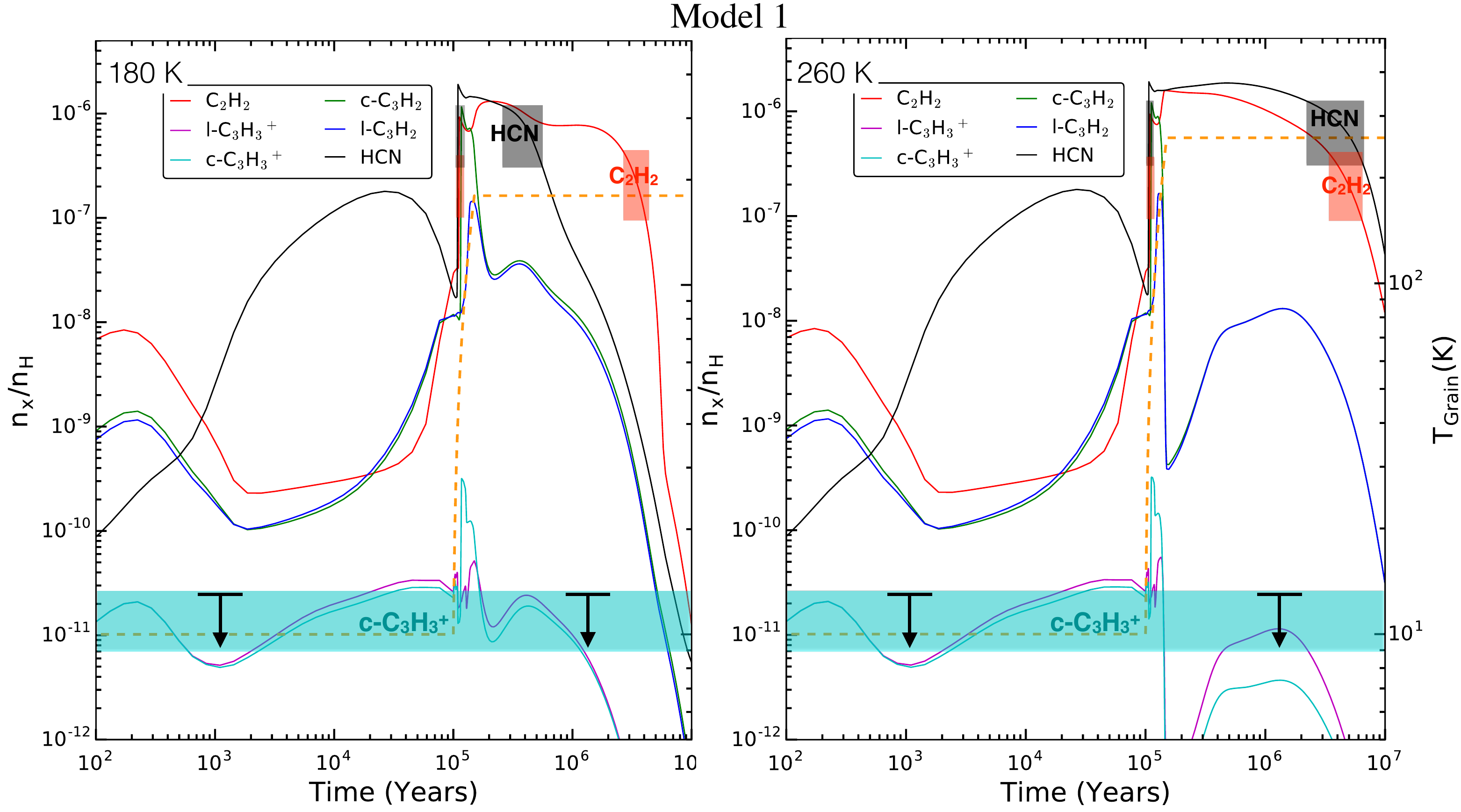}
\caption{Abundance predictions (solid lines) from our model simulating hot core chemistry. The color bands indicate the range of values consistent with SOFIA/EXES measurements. Dashed orange line is the gas/dust temperature -- scale shown on the opposite y-axis. See text for model description.}
\label{model1}
\end{figure*}

Recent observations conclude that the Orion hot core may not be internally heated as would be the case for a typical hot core with an embedded protostar \citep{zapata2011, goddi2011, bally2017, wright2017}. These studies find that, unlike other hot cores, the hot molecular gas emission from the Orion hot core is not associated with any self-luminous submillimeter, radio, or infrared source.  Alternative explanation for heating the hot core is by a group of strong near- and mid-infrared compact sources located very close to the core (e.g., IRc2). However, \citet{zapata2011} explains that sensitive VLA radio observations made by \citet{menten95} revealed that most of these IR luminous sources in the Orion-KL nebula are not self-luminous but rather show reprocessed emission escaping through inhomogeneities in the dense material. 

It has been proposed that the Orion hot core could be a result of a pre-existing density enhancement heated from the outside by the explosive event that took place 500 yrs ago, behind the Orion nebula. This event resulted in high-velocity streamers (10 to $>100$ \kms) seen in CO \citet[e.g.,][]{zapata2009, bally2011, bally2017}, FeII and \htwo\ \citep{bally2015}. However, as \citep{zapata2011} explains, the CO filamentary streamers are absent from the area behind the Orion KL hot core (i.e. behind relative to the outflow center), suggesting that a dense material might have impeded the expansion of these filaments. Shocks resulting form the explosive event then created the ``hot core". If this is the case then the Orion hot core could be as young as 500 yrs. A low velocity, C-type shock can heat the hot core and drive the chemistry \citep[e.g.,][]{harada2010}, which is consistent with the large abundances in complex organic molecules seen in various observations towards Orion-KL \citep{sutton85, crockett14, feng2016}. 

Even though the Orion hot core may not be heated internally, it is still interesting to compare our observations with a typical hot core chemistry model to understand how it might differ from the above scenario. 
A typical model for hot core chemistry includes three phases \citep[e.g.,][]{garrod2006, garrod2008}. In the initial stage, the cold core (T$\sim$10 K) undergoes an isothermal collapse during which the accretion of atoms and simple molecules from the gas phase onto dust grains promotes a surface chemistry that results in the build up of icy mantles upon the grains. Later, the formation of a nearby protostar quickly (in $\sim 5 \times 10^4$ years) warms the gas and dust (T: 100--300~K), sublimating the grain mantle material into the gas phase. In the final hot core phase, the evaporated molecules subsequently drive a rapid gas-phase chemistry in dense, warm gas. 

We ran gas-grain chemical models containing a network of 925 species and 12000 reactions that estimate the abundances of molecules as a function of time and physical conditions. For a prototypical hot core with an embedded young stellar object, the chemical simulations are run in a warm up mode, in which the temperature increases as the source of the infrared radiation turns on and heats up \citep{garrod2006}. The gas and grain chemistry is treated via the standard rate equation approach \citep{hasegawa92}.
We used a warm-up model as described in \citet{garrod2008}, which is suitable for hot molecular cores. The chemical network used is sensitive up to 800 K and described in \citet{acharyya2015a, acharyya2015b}.
The major features of the model used are as follows:

\begin{itemize}
\item The so-called classical dust grains have a size of 0.1 $\mu$m with a surface site density $n_{\rm s}$ = 1.5 $\times$ 
10$^{15}$ cm$^{-2}$, leading to about 10$^6$ binding sites of adsorption per grain.

\item Standard low-metal elemental abundances, initially in the form of gaseous atoms --- except for molecular hydrogen -- are used. 
Elements having ionization potentials lower than 13.6 eV are in the form of singly charged positive ions, i.e., C$^+$, Fe$^+$, 
Na$^+$, Mg$^+$, S$^+$, Si$^+$, and Cl$^+$.

\item Physical parameters other than gas and grain temperatures remain constant and homogeneous throughout the chemical evolution 
with a proton density $n_{\rm H}$ = $2 n(\rm H_{2}) + n(\rm H) $ = 10$^5$ cm$^{-3}$ , visual extinction A$_{\rm V}$= 10 mag and 
standard cosmic ray radiation flux of 10$^{-17}$ s$^{-1}$.

\item Binding energy for desorption and hopping is taken from \citet{garrod2008}. The diffusion-to-binding energy ratio
is a key parameter for the study of grain surface chemistry although poorly constrained. \citet{katz99} found this ratio 
to be $\sim$ 0.8 for hydrogen by fitting laboratory data. Recently, \citet{he2016b} found binding energy to be coverage 
dependent for a few species. In the absence of detailed laboratory data for most of the species, several values 
are used in the literature; $0.3$ \citep{hasegawa92}, $0.5$ \citep{garrod2006,garrod2008, acharyya2011}, and a time dependent 
value \citep{garrod2011}. We used a value of 0.5 for the diffusion-to-binding energy ratio in accordance with \cite{garrod2008}.

\item A sticking coefficient is usually between 0 and 1 for astrochemical modeling. However, a recent formulation as a function of temperature is provided by \cite{he2016a}, based on laboratory measurement. Since temperature changes are a critical part of our models, we used a sticking coefficient based on this paper.
\end{itemize}

Initially both the gas and grain temperatures are kept constant at 10 K for 10$^5$ years; then the temperature is linearly increased to 180 K or 260 K over $5 \times 10^{4}$ yrs. The hot core chemistry was then followed for 10$^7$ years. Figure~\ref{model1} shows the time variation of selected species in this model for 180~K (left panel) and 260~K (right panel).\\

The dominant reactions leading to the formation of \acet\ in the gas phase and on grains are presented in Figure \ref{opr} (left panel). The best pathways to form HCN during the cold phase are 

\begin{eqnarray}\nonumber
N + CH_{2}  \rightarrow  HCN + H\\
\nonumber
HCNH^{+} + e  \rightarrow  HCN + H\\
\nonumber
C_{2}N + H  \rightarrow  HCN + H\\
\nonumber
HNC + C  \rightarrow  HCN + C\\
\nonumber
\end{eqnarray}

Once the temperature starts to increase during warm-up, thermal desorption becomes dominant until the sublimation of all the HCN from grains is complete. After that there are three dominant gas phase formation pathways: 
\begin{eqnarray}
\nonumber
NH_{3} + HCNH^{+} \rightarrow HCN + NH_{4}^{+} \\
\nonumber
 HCNH^{+} + e \rightarrow HCN + H\\
 \nonumber
CN + H_{2} \rightarrow HCN + H\\
\nonumber
\end{eqnarray}

In Figure \ref{model1}, we can compare observed and model abundances for HCN and \acet\ as a function of hot core age. The shaded regions show the range of observed abundances where they agree with the model predictions. The observed abundances can be produced only in the hot core phase (or after the warm-up phase). This is true for both 180\,K and 260\,K hot core models. At 180\,K, HCN and \acet\ abundances barely overlap, whereas at 260~K, they overlap over roughly $(2 - 5) \times 10^{6}$ years.  This suggests the two temperature limits generally encompass the range where observed abundances from the two species will overlap. Note that the observed abundances also fortuitously match the model at the on-set of the warm-up phase when there is a steep rise in the gas-phase molecular abundances. The observations constrain the age of the hot core to be roughly $10^{6}$ years. Even though, our observed abundances agree with the hot core model presented here, the timescale needed to produce these abundances is too long to be consistent with the 500 yr timescale of the explosive event.  

Another scenario is presented by \citet{goddi2011}.  They observed seven metastable inversion transitions of ammonia (\amon): (J, K) = (6,6) to (12,12) that have energy level of the ground state $\sim1500$\,K, allowing them to trace hot molecular gas at high spatial resolution in the vicinity of Source I and the hot core. Their study also favors an external heating source for the Orion hot core -- a low-velocity (10 - 50 \kms) shock hypothesis based on the observed offset of hot \amon\ emission peaks from known protostellar sources, high kinetic temperatures, and enhanced ortho to para ratio in \amon. However, they believe that the impact of an outflow from Source I (20 \kms) in combination with proper motion of 12 \kms\ is responsible for heating the core rather than the shocks generated from the recent explosive event. Very recent ALMA 349 GHz observations of dust and molecular emission by \citet{wright2017} argue that either shock scenario could be responsible for exciting the hot core. These ALMA observations reveal a remarkable molecular ring, $\sim 2$\arcsec\ south of Source I, with a diameter $\sim$600~AU. The ring was seen in high excitation transitions of HC$_3$N, HCN v2=1, and SO$_2$ and is thought to originate from an impact between the ejecta from the BN/Source I explosion and a dense dust clump. In conclusion, detailed modeling that includes shock physics as well as dust/gas-phase chemistry, will be required to appropriately discriminate between various shock scenarios proposed for the excitation of the Orion hot core.

%{\bf Question: Can a low-velocity shock (either from an explosion or an outflow from source I) produce the high abundances seen in \acet\ and HCN? Recent observation strongly indicate that there was a pre-existing density enhancement that was impacted by shocks from an explosive event or an outflow. Can a C-type low velocity shock sublimate icy mantles and release C2H2 in gas-phase like assumed in a hot core chemistry and match our large observed abundances? Also, can a C-type shock do this in 500 yrs?}

\subsection{Non-Equilibrium Ortho to Para Ratio in Low Velocity Shocks}
\begin{figure*}
\center
\includegraphics[scale=0.36]{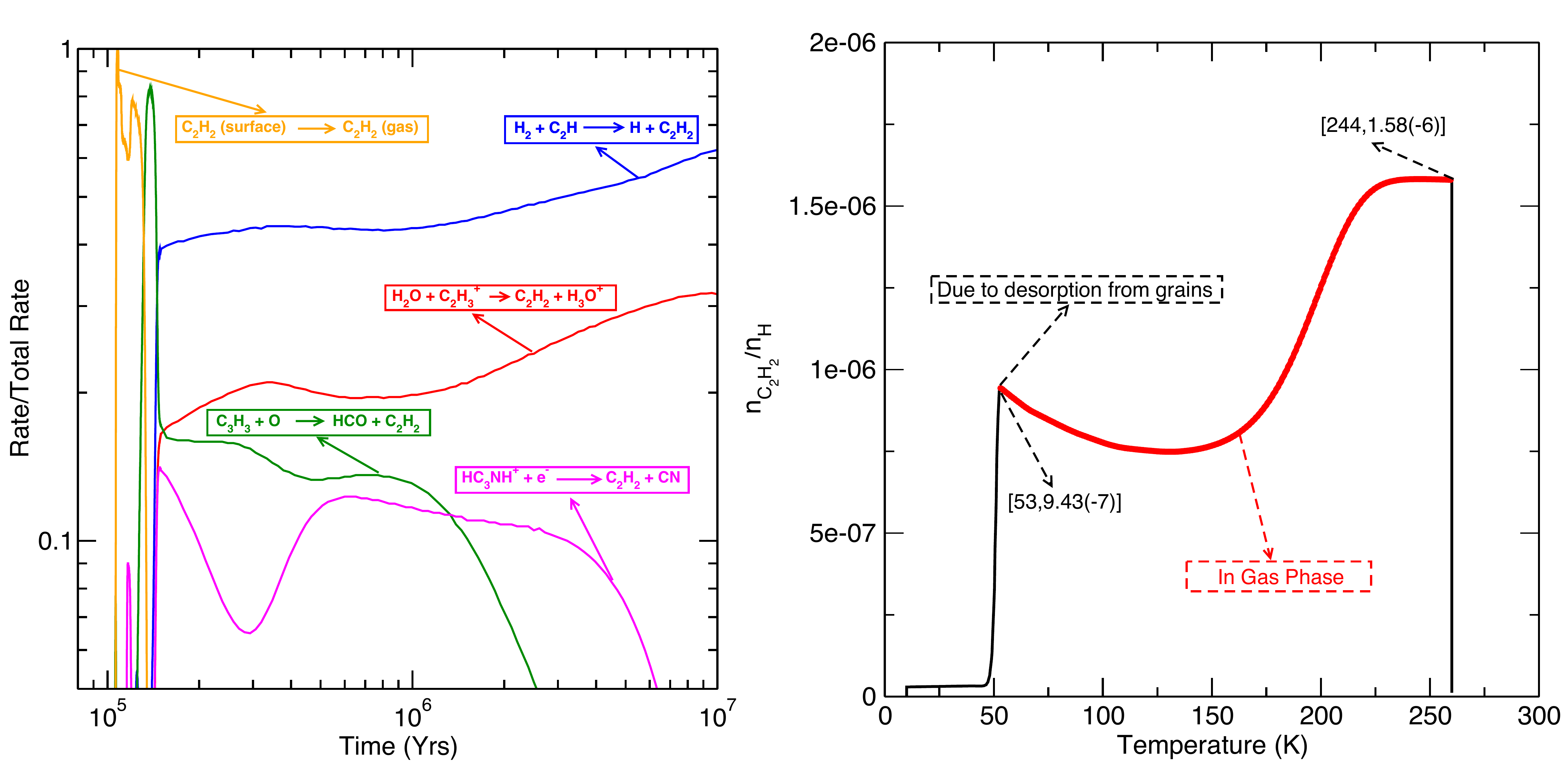}
\caption{(Left) Dominant formation reactions for \acet\ on grain surfaces and in the gas phase, as a function of time during and post warm-up phase. (Right) The abundance is significantly increased via gas phase reactions (red part) once the ambient temperature is higher than the \acet\ desorption temperature. At high temperatures, the \acet\ residence time is very low, therefore, the OPR can not be changed by the grains.}
\label{opr}
\end{figure*}
Any species with two identical hydrogen nuclei exists in two nuclear spin configurations, labeled ortho and para. Since spontaneous conversions between the different spin symmetries are extremely slow due to very weak magnetic interactions between intramolecular nuclear spins \citep{pachucki2008}, ortho and para \acet\ can often be distinguished as two different molecules. In LTE at high temperatures, the OPR is 3, reflecting the statistical ratio of spin states. Since there is no study on \acet\ OPR formation on grains yet,  we assume that the \acet\ OPR will follow similar trends as seen for the \htwo\ OPR in shock models, i.e., both \htwo\ and \acet\ have a high temperature OPR limit of 3, which decreases with decreasing temperature. There are several possible ortho-to-para conversion processes, which can occur in the gas phase: (1) proton exchange of neutral species, such as  \htwo\, with abundant protonated ions such as H$^{+}$, H$_{3}^{+}$ and (2) atomic H-exchange with ionic or radical species. In the solid phase, ortho-para interconversions can occur through interaction of the nuclear spins of the di-hydrogenated species with paramagnetic impurities or imperfections on dust grain surfaces (Fukutani \& Sugimoto 2013 and references therein). At low temperatures (T: 10 -- 15 K), the ortho to para conversion for \htwo\ on grains is thought to be much faster than in the gas phase \citep{bovino2017}, reducing the time needed for \htwo\ OPR to reach its low temperature equilibrium value. In C-type shocks ($V_\mathrm{s}$: 10 -- 30 \kms), the densities of H$^{+}$ and H$_{3}^{+}$ are generally found to be low \citep{timmermann98, wilgenbus2000}. Hence, process (1) can convert only a small fraction of para-\htwo\ into ortho-\htwo\ during the passage of the shock. According to \citet{timmermann98}, proton exchange reactions are considered the principal process for shock speeds $V_\mathrm{s} \lesssim 15$ \kms. At higher shock speeds, the OPR reaches its equilibrium value quickly through reactive H-\htwo\ collisions because of higher kinetic temperatures and the atomic hydrogen abundance reached in the shocked gas. 

Observationally, abnormally low OPRs have been found for \htwo\ in various environments: towards molecular clouds near the Galactic center \citep{rodriguez2000}, reflection nebula NGC 7023 \citep{fuente99} and in shocked molecular gas \citep{neufeld2006}. All these studies see a similar `zig-zag' pattern in the rotational diagram or the offset in the ortho and para rotational ladder that we find for \acet\ in this work (Figures \ref{comp} \& \ref{rotdia}). C-type shock models by \citet{timmermann98} and \citet{wilgenbus2000} were used to explain low OPR observed in \htwo. For the Orion hot core, the observations show enhancements in the \amon\ OPR \citep{goddi2011}, but declines in the \acet\ OPR. 

We propose here a plausible scenario: the abnormally low \acet\ OPR observed could be a remnant from an earlier, colder phase, before the density enhancement was impacted by shocks 500 yrs ago. This will imply a lower thermal equilibrium OPR either in the gas phase or frozen into dust-grain surfaces before being released into the gas phase by shocks. For a pre-shock para-enriched \htwo\ gas with an OPR lower than the high temperature statistical limit of 3, magnetohydrodynamics (MHD) shock modeling studies have indeed shown that if the shock velocity is below a threshold value of $\sim$ 20-25 \kms, the post-shock \htwo\ OPR does not reach its high-temperature LTE value, because the OPR thermalization timescale is longer than the shock timescale \citep{timmermann98}. As shown in \citet{faure2013}, the LTE curves for \htwo\ and NH$_3$ as a function of temperature for colder, pre-shock temperatures of 10 -- 50 K, have opposite trends. The \htwo\ LTE OPR decreases with the decreasing temperature, whereas the NH$_3$ LTE OPR increases. This is consistent with our observations of \acet\ and NH$_3$ observations by \citet{goddi2011}. Thus, one can hypothesize that not enough time has passed for the OPR of \amon\ and \acet\ to reach their equilibrium values at post-shock high temperatures, and that we are probing the OPR of the earlier, colder phase of the core before it was impacted by shocks associated with either the explosion or an outflow from source I.

Higher than normal ambient temperatures ($\sim 50$~K) are quite possible in this high mass star forming region with strong IR emission. If the pre-shock temperature was below the \acet\ desorption temperature ($\sim 50$~K), then a large quantity of \acet\ might also lie on grain surfaces, where ortho to para conversion can take place. However, if the temperature is above the desorption temperature then \acet\ should be mainly in the gas phase and its abundance will increase via
gas phase reactions allowing ortho to para conversion to take place there (see Figure \ref{opr} right panel). However, these gas-phase conversion processes might be very slow compared to the age of the probed gas -- not sufficiently rapid to thermalize the \acet\ OPR to the gas temperature. A low \acet\ OPR could thus reflect a para-enriched \htwo\ gas, provided the memory of chemical processes can be propagated and preserved in the molecular level population distribution following the nuclear spin selection rules (Oka 2004; Hugo et al. 2009; Faure et al. 2013; Le Gal et al. 2014) for the dominant gas-phase formation pathway C$_2$H + H$_2 \rightarrow $ C$_2$H$_2$ + H. In a nutshell, the mechanism for ortho to para conversion of \acet\ should greatly depend on the ambient temperature before the material is shocked. 

\subsection{Search for c-\xmol}

This project began as a search for c-\xmol, which is considered the most important precursor in the formation of the interstellar organic ring molecule Cyclopropenylidene (c-$\mathrm{C}_{3}\mathrm{H}_{2}$). It is known that c-$\mathrm{C}_{3}\mathrm{H}_{2}$ is ubiqutuous in the interstellar medium (ISM) -- being found in giant molecular clouds, dark dust clouds, diffuse cloud regions, circumstellar envelopes, and other galaxies \citep[e.g.,][]{madden89}, but c-\xmol\ has never been detected in the ISM. Along with PAHs, c-$\mathrm{C}_{3}\mathrm{H}_{2}$ is of particular interest because of its possible relationship both to more complex aromatic compounds in the ISM as well as terrestrial biochemistry. The gas-phase chemistry models \citep{harada2010, mookerjea2012} yield a wide range in c-\xmol/c-$\mathrm{C}_{3}\mathrm{H}_{2}$ abundance ratios of 0.001 to 1.0. Understanding how c-$\mathrm{C}_{3}\mathrm{H}_{2}$ forms requires measurement of the c-\xmol/c-$\mathrm{C}_{3}\mathrm{H}_{2}$ abundance ratio. In fact, if the c-\xmol/c-$\mathrm{C}_{3}\mathrm{H}_{2}$ abundance ratio is sufficiently below predicted values, the chemical networks would require significant revision. For this reason, there has been interest in detecting c-\xmol\ in astrophysical environments for nearly 25 years \citep[][and references therein]{lee89}. The search for this molecule has been made difficult due to lack of experimental data. Also, c-\xmol\ has no permanent dipole moment so it is only accessible in the MIR, where its rovibrational transitions lie. 
Portions of the \xmol\ spectrum have been measured in the lab but with only moderate
resolution \citep{ricks2010}, or in high-resolution but for a band outside the range of the EXES instrument \citep{zhao2014}. \citet{huang2011a} and \citet{huang2011b} have reported on very accurate quartic force field and spectroscopic constants from  {\it ab initio} calculations for \xmol. Predictions from those calculations motivated our search for this molecule with SOFIA. 

Our EXES observations were optimized to the $\nu_{7}$ band of \xmol, predicted to be around 13.21 \mic, with theoretical uncertainties that can shift this band position by +/- 3 \cmone\ ($\sim$0.06 \mic). We measure a 3$\sigma$ upper limit on the column density of $2 \times 10^{13}$ \cmtwo, assuming that the $\nu_{7}$ transitions lie within our observed bandpass. This upper limit is consistent with the hot core model shown in Figure \ref{model1}. Our next set of scheduled observations with SOFIA will continue the search for this molecule with longer integration times and wider wavelength coverage.  There is no study on how c-\xmol\ chemistry would evolve in low-velocity shocks.  

\section{Conclusions}
We present mid infrared, high spectral resolution, observations towards Orion IRc2, sampling the chemical and physical conditions of the Orion hot molecular core. These observations yielded high S/N, continuous coverage of the R-branch \acet\ lines from J=9-8 to J=18-17, including both ortho and para species. Eight of these rovibrational transitions are newly reported detections. The detection of the isotopologue, \acetiso, enabled {\bf a} direct measurement of the \cratio\ isotopic ratio for the Orion hot core of $14 \pm 1$ and an estimated maximum (3$\sigma$) value of 21. The upper limit is consistent with the previously reported values in the literature. We unambiguously detect an offset between the ortho and para \acet, implying that they are coming from gas at two different kinetic temperatures of 226 K and 164 K, respectively. This combined with the significant differences in their \vlsr\ and line widths also suggest that the two species may not be exactly co-located. We are not aware of any published models that can explain physical/chemical processes leading to such a spatial ortho-para separation. Future effort on shock induced chemistry models may shed more light on this result. In the laboratory, ortho-para separations have been achieved by the use of heterogeneous electric fields with a J-dependent interaction, and by different  rates of ortho and  para adsorption onto a number of solids such as carbon \citep[e.g.,][]{tikhonov2002}. We measure the ortho to para ratio of $1.7 \pm 0.1$, which is low compared to its the high temperature LTE value of 3. We propose that this low OPR ratio could be a remnant from an earlier, colder phase, before the density enhancement (now the hot core) was impacted by shocks 500 yrs ago, based on the relatively longer timescales for thermalization in shocks.

\acknowledgements
We thank our referee for constructive comments that have significantly strengthened the paper. We are grateful to John Lacy for sharing his TEXES observations and findings with us that led to a better understanding of our EXES data.
We thank Prof. Joel Bowman (Emory University) for useful discussions. 
This work was supported by the SOFIA observing grant, SOF 03-0126.  Dr. Rangwala acknowledges support from the NASA Ames/BAER institute cooperative agreement NNX14AR61A. Dr. Huang acknowledges support from the NASA/SETI Institute Cooperative Agreement NNX15AF45A. Prof. E. Herbst thanks the NSF for support of his program in astrochemistry through  grant AST 1514844.

\bibliographystyle{apj}
%%\bibliography{astrochem}

\end{document}